\documentclass[pre,superscriptaddress,twocolumn,amsmath,amssymb,floatfix]{revtex4}

\usepackage{graphicx}

\begin{document}

\title{Generalized friendship paradox in networks with tunable degree-attribute correlation} 
\author{Hang-Hyun Jo}
\altaffiliation[Present address: ]{BK21plus Physics Division and Department of Physics, Pohang University of Science and Technology, Pohang 790-784, Republic of Korea}
\affiliation{BECS, Aalto University School of Science, P.O. Box 12200, Espoo, Finland}
\author{Young-Ho Eom}
\affiliation{Laboratoire de Physique Th\'eorique du CNRS, IRSAMC, Universit\'e de Toulouse, UPS, F-31062 Toulouse, France}

\date{\today}

\begin{abstract}
One of interesting phenomena due to topological heterogeneities in complex networks is the friendship paradox: Your friends have on average more friends than you do. Recently, this paradox has been generalized for arbitrary node attributes, called generalized friendship paradox (GFP). The origin of GFP at the network level has been shown to be rooted in positive correlations between degrees and attributes. However, how the GFP holds for individual nodes needs to be understood in more detail. For this, we first analyze a solvable model to characterize the paradox holding probability of nodes for the uncorrelated case. Then we numerically study the correlated model of networks with tunable degree-degree and degree-attribute correlations. In contrast to the network level, we find at the individual level that the relevance of degree-attribute correlation to the paradox holding probability may depend on whether the network is assortative or dissortative. These findings help us to understand the interplay between topological structure and node attributes in complex networks.
\end{abstract}



\maketitle

\section{Introduction}

Human societies have been successfully described within the framework of complex networks, where nodes and links denote individuals and their dyadic relationships, respectively~\cite{Albert2002,Castellano2009,Borgatti2009,Lazer2009,Holme2012}. As individuals are embedded in social networks, their positions in such networks strongly influence their behaviors~\cite{Borgatti2009} as well as self-evaluations~\cite{Zuckerman2001} and subjective well-being~\cite{Kross2013}. In particular, the comparison to friends, colleagues, and peers enables individuals to adopt and transmit opinion, information, and technologies~\cite{Castellano2009,Easley2010,Pinheiro2014}, e.g., for the competitiveness~\cite{Garcia2013}. Thus understanding positional differences between individuals is crucial to understand the emergent collective dynamics at the community or societal level~\cite{Buchanan2007}. 

Topological structures of social networks have been known to be heterogeneous, characterized by broad distributions of the number of neighbors or degree~\cite{Barabasi1999}, assortative mixing~\cite{Newman2002}, and community structure~\cite{Fortunato2010}. One of interesting phenomena due to topological heterogeneities is the friendship paradox (FP). The FP states that your friends have on average more friends than you do~\cite{Feld1991}. The paradox has been shown to hold in both offline and online social networks~\cite{Feld1991,Ugander2011,Hodas2013,Eom2014,Kooti2014,Grund2014}. Examples include friendship networks of middle and high school students~\cite{Feld1991,Grund2014} and of university students~\cite{Zuckerman2001}, scientific collaboration networks~\cite{Eom2014}, and Facebook and Twitter user networks~\cite{Ugander2011,Hodas2013,Kooti2014}. The paradox can be understood as a sampling bias in which individuals having more friends are more likely to be observed by their friends. This bias has important implications for the dynamical processes on social networks, e.g., for efficient immunization~\cite{Cohen2003} and for early detection of contagious outbreaks~\cite{Christakis2010,Garcia-Herranz2014} or of natural disasters~\cite{Kryvasheyeu2014}. The paradox implies that your friends and neighbors tend to occupy more important or central positions in social networks than you do.

The importance or centrality of individuals is not determined only by their topological positions in networks, but also influenced by their attributes. Individuals can be described by various attributes like gender, age, cultural preferences, and genetic information~\cite{Park2007,Fowler2008}. This requires us to study the interplay between topological structure and node attributes of social networks. The friendship paradox has been also considered for arbitrary node attributes~\cite{Hodas2013,Eom2014,Kooti2014}, which is called generalized friendship paradox (GFP)~\cite{Eom2014}. Note that if the degree of node is considered as the attribute, the GFP reduces to the FP. 

The GFP can be formulated at the individual and network levels. The GFP holds for a network if the average attribute of nodes in the network is smaller than the average attribute of their neighbors. The GFP holds for a node if the node has lower attribute than the average attribute of its neighbors. The GFP at both levels has been observed in the coauthorship networks~\cite{Eom2014}. While the GFP at the network level accounts for the average behavior of the network, the GFP at the individual level can provide more detailed understanding of the centrality of individuals, and of their subjective evaluations of attributes. It is obvious that these individual properties cannot be fully revealed in the network level analysis, especially when the individuals are heterogeneous, e.g., in terms of broad distributions of degree and attribute.

The origin of the GFP at the network level has been clearly shown to be rooted in positive degree-attribute correlations~\cite{Eom2014}. In other words, high attribute individuals are more likely to be observed by their friends as high attribute individuals have more friends. However, the role of degree-attribute correlations at the individual level is far from being fully understood. In order to investigate the role of various correlations for the GFP at the individual level, we first analyze a solvable model to characterize the paradox holding probability of nodes for the uncorrelated case. Then we numerically study the correlated model of networks with tunable degree-degree and degree-attribute correlations. By calculating the paradox holding probabilities for the entire range of correlations, we show that the relevance of degree-attribute correlation to the paradox holding probability may depend on whether the network is assortative or dissortative. This result is compared to the GFP at the network level. Finally, we conclude the paper by summarizing the results.

\section{Generalized friendship paradox}

\subsection{Network level}

The generalized friendship paradox (GFP) holds for a network if the average attribute of nodes in the network is smaller than the average attribute of their neighbors. For a network of $N$ nodes, let us denote a degree and an attribute of node $i$ as $k_i$ and $x_i$, respectively. The average degree and average attribute are $\langle k\rangle=N^{-1}\sum_{i=1}^N k_i$ and $\langle x\rangle=N^{-1}\sum_{i=1}^N x_i$. The average attribute of neighbors $\langle x\rangle_{nn}$ is obtained as
\begin{equation}
  \langle x\rangle_{nn}=\frac{\sum_{i=1}^N k_ix_i}{\sum_{i=1}^N k_i},
\end{equation}
where a node $i$ with degree $k_i$ has been counted $k_i$ times by its neighbors. Then the GFP holds for a network if the following condition is satisfied:
\begin{equation}
  \langle x\rangle<\langle x\rangle_{nn}.
\end{equation}
By the straightforward calculation, one gets
\begin{equation}
  \label{eq:networkGFP}
  \langle x\rangle_{nn}-\langle x\rangle=\frac{\rho_{kx}\sigma_k\sigma_x}{\langle k\rangle},
\end{equation}
where the degree-attribute correlation is given by
\begin{equation}
  \rho_{kx}=\frac{1}{N}\sum_{i=1}^N\frac{(k_i-\langle k\rangle)(x_i-\langle x\rangle)}
  {\sigma_k\sigma_x}.
\end{equation}
Since standard deviations of degree and attribute, i.e., $\sigma_k$ and $\sigma_x$, are positive in any non-trivial cases, the positive $\rho_{kx}$ leads to the GFP at the network level. Thus, the origin of GFP at the network level is rooted in positive correlation between degree and attribute~\cite{Eom2014}. The GFP at the network level has been observed in the coauthorship networks of Physical Review journals (PR) and of Google Scholar profiles (GS) for several attributes such as the number of publications by each author~\cite{Eom2014}. In addition, the negative $\rho_{kx}$ can lead to the opposite tendency, implying that your friends have on average lower attribute than you do. This can be called anti-GFP.

\subsection{Individual level: Uncorrelated solvable model}

In order to investigate the GFP at the individual level, we study an uncorrelated solvable model. The GFP holds for a node $i$ if the node has lower attribute than the average attribute of its neighbors, precisely if the following condition is satisfied:
\begin{equation}
  \label{eq:individualGFP}
  x_i<\frac{1}{k_i}\sum_{j\in\Lambda_i}x_j,
\end{equation}
where $\Lambda_i$ denotes the set of $i$'s neighbors. The probability of satisfying Eq.~(\ref{eq:individualGFP}) or \emph{paradox holding probability} may be interpreted as the degree of self-evaluation of the node when compared to its neighbors. We assume no correlation between attributes of neighboring nodes, implying that the degrees of neighbors are entirely irrelevant to the probability. Then one gets the paradox holding probability of a node with degree $k$ and attribute $x$ as
\begin{eqnarray}
  h_k(x)&\equiv&\Pr\left(\frac{1}{k}\sum_{j=1}^k x_j>x\right)\\
  &=&\prod_{j=1}^k \int_0^\infty dx_j P(x_j)\theta\left(\frac{1}{k}\sum_{j=1}^k x_j-x\right),
\end{eqnarray}
where $\theta(\cdot)$ is a Heaviside step function. The distribution of $x$ has been denoted by $P(x)$ with $x\geq 0$. In general $x$ can have negative values, which will be considered in the next Subsection. By taking the Laplace transform with respect to $x$, we get
\begin{eqnarray}
  \tilde h_k(s)=\frac{1}{s}\left[1-\tilde P\left(\frac{s}{k}\right)^k\right],
\end{eqnarray}
where $\tilde P(s)$ is the Laplace transform of $P(x)$. Then, the paradox holding probability $h_k(x)$ can be obtained by taking the inverse Laplace transform of $\tilde h_k(s)$ analytically or numerically if necessary.

\begin{figure}[!t]
    \includegraphics[width=.95\columnwidth]{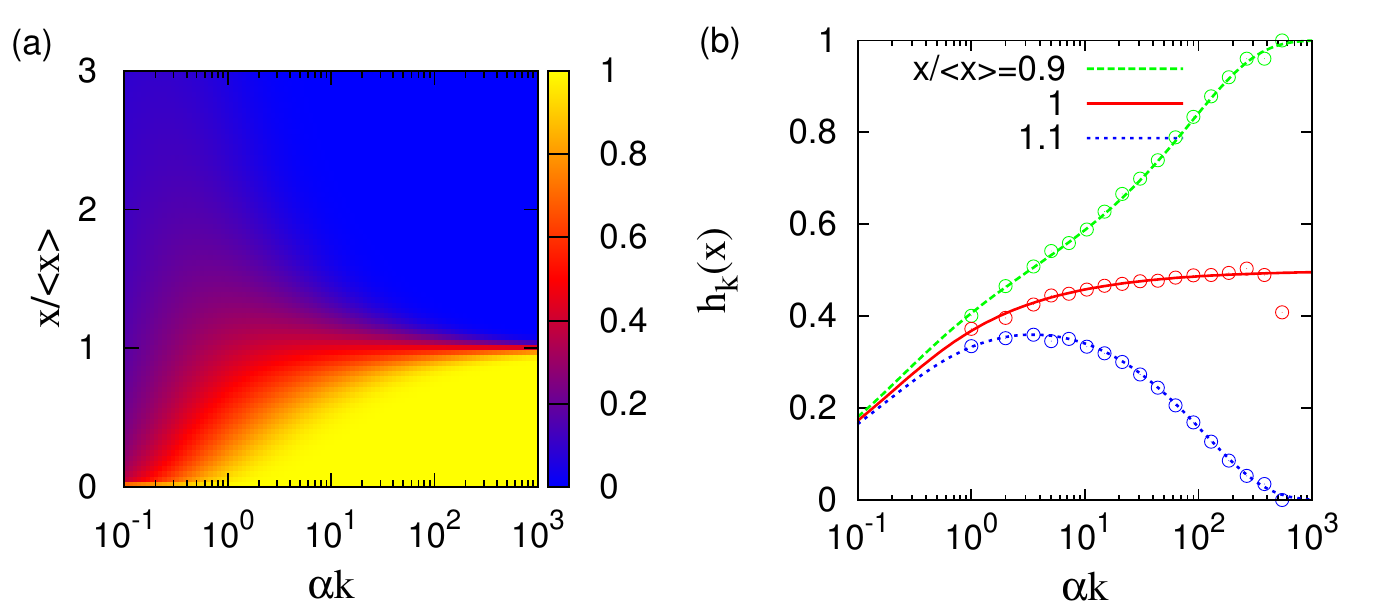}
    \caption{(Color online) Analytic results of the uncorrelated model with gamma distributions for $x$ and $k$ in Eq.~(\ref{eq:gammaDistr}). (a) Heat map of the paradox holding probability $h_k(x)$ in Eq.~(\ref{eq:Pkx_gamma}) as a function of $\alpha k$ and $x/\langle x\rangle$. (b) $h_k(x)$ as a function of $\alpha k$ for values of $x/\langle x\rangle=0.9$, $1$, and $1.1$ (curves), which are compared to the numerical results (circles) from the uncorrelated network of size $N=10^5$ and of $\langle x\rangle=50$ using the same gamma distribution in Eq.~(\ref{eq:gammaDistr}).}
    \label{fig:gamma_uncorrel}
\end{figure}

For the solvable yet broadly distributed case, we consider the gamma distribution for $x$, i.e.,
\begin{eqnarray}
  \label{eq:gammaDistr}
  P(x)=\frac{x^{\alpha-1}e^{-x/\beta}}{\beta^\alpha \Gamma(\alpha)},
\end{eqnarray}
where $\alpha, \beta>0$ and the mean of $x$ is $\langle x\rangle=\alpha\beta$. Since $\tilde P(s)=(\beta s+1)^{-\alpha}$, one gets
\begin{eqnarray}
  \label{eq:Pkx_gamma}
  h_k(x)=\frac{\Gamma(\alpha k,\alpha k \frac{x}{\langle x\rangle})}{\Gamma(\alpha k)}.
\end{eqnarray}
Here $\Gamma(s,z)=\int_z^\infty t^{s-1}e^{-t}dt$ denotes the upper incomplete gamma function. The heat map of $h_k(x)$ as a function of $\alpha k$ and $x/\langle x\rangle$ is depicted in Fig.~\ref{fig:gamma_uncorrel}(a).

For any given $k$, it is obvious that $h_k(0)=1$ and $h_k(\infty)=0$, and that $h_k(x)$ is a decreasing function of $x$. For a given $x$, one can study the $k$-dependent behavior of $h_k(x)$. In case of $k=1$, $h_1(x)$ is the probability of drawing one number larger than $x$ from $P(x)$, which we denote $f_x\equiv \int_x^\infty P(x')dx'$. The value of $h_2(x)$ is upper bounded by the probability that when two numbers are drawn from $P(x)$, both numbers are not smaller than $x$, i.e., $h_2(x)\leq 1-(1-f_x)^2$. Even when one neighbor has an attribute less than $x$ and the other has an attribute more than $x$, it is likely that the average of them exceeds $x$ due to the broadness of $P(x)$. Thus, we approximate as $h_2(x)\approx 1-(1-f_x)^2$, which is then generalized to $h_k(x)\approx 1-(1-f_x)^k$. This argument accounts for the $k$-dependent increasing behavior for small $\alpha k$ in the solution of Eq.~(\ref{eq:Pkx_gamma}). It could imply that having more friends may lead to the lower self-evaluation to some extent. However, for sufficiently large $k$, the average of attributes of neighbors converges to $\langle x\rangle$. Hence, when the given $x$ is smaller (larger) than $\langle x\rangle$, $h_k(x)$ approaches $1$ ($0$) as $k$ increases. In case of $x=\langle x\rangle$, $h_k(x)$ approaches $1/2$ as $k$ increases. Note that only when $x>\langle x\rangle$, $h_k(x)$ increases and then decreases according to $k$. Such nontrivial behavior emerges even in the uncorrelated case.

Next, in order to study the FP in the uncorrelated setup, one needs to solve the following equation:
\begin{eqnarray}
  h^{\rm FP}_k&\equiv&\Pr\left(\frac{1}{k}\sum_{j=1}^k k_j>k\right)\\
  &=&\sum_{\{k_j\}} \prod_{j=1}^k P(k_j)\theta\left(\frac{1}{k}\sum_{j=1}^k k_j-k\right),
\end{eqnarray}
where $P(k)$ denotes the degree distribution. As there is no general solution to our knowledge, the FP will be numerically studied in the next Subsection.

\subsection{Individual level: Correlated network model}

We numerically study more general cases, including the uncorrelated model, by generating networks with tunable degree-degree and degree-attribute correlations. Following the configuration model~\cite{Catanzaro2005}, we generate the degree sequence, $\{k_i\}$ for nodes $i=1,\cdots,N$, where each degree is independently drawn from $P(k)$ with minimum degree as $k_{\rm min}=1$. Each node has $k_i$ ``stubs'' or half links. A pair of nodes are randomly selected and a link is established between them if both nodes have residual stubs and if there is no link between them. This process is repeated until when no stubs remain. In principle, the generated network has no degree-degree correlations. The degree-degree correlation can be characterized by the assortativity coefficient~\cite{Newman2002}
\begin{equation}
  \label{eq:assort}
  r_{kk} = \frac{L\sum_l k_lk'_l-[\sum_l\frac{1}{2}(k_l+k'_l)]^2} {L\sum_l\frac{1}{2}({k_l}^2+{k'_l}^2) -[\sum_l\frac{1}{2}(k_l+k'_l)]^2},
\end{equation}
where $k_l$ and $k'_l$ denote degrees of nodes of the $l$th link with $l=1,\cdots,L$, and $L$ is the total number of links in the network. The value of $r_{kk}$ ranges from $-1$ to $1$, and it quantifies the tendency of large degree nodes being connected to other large degree nodes. A network with the maximal $r_{kk}$ can be implemented, e.g., by constructing $k$-cliques or complete subgraph with $k$ nodes. The minimal $r_{kk}$ can be found in the star-like network structure, where hubs are connected to dangling nodes. For preparing the network with a desired value of $r_{kk}$, we rewire links as following~\cite{Maslov2002}: Two links are randomly selected, e.g., a link between nodes $i$ and $j$ and a link between nodes $i'$ and $j'$. These nodes are rewired to links between $i$ and $i'$ and between $j$ and $j'$, only when the value of $r_{kk}$ gets closer to the desired value. This rewiring is repeated until when the desired value of $r_{kk}$ is reached.

For the tunable degree-attribute correlation, denoted by $\rho_{kx}$, we adopt the method used in~\cite{Eom2014}. For a given degree sequence, the attribute of a node $i$ is assigned as
\begin{equation}
  \label{eq:assign_x}
  x_i=\rho k_i+\sqrt{1-\rho^2}k_j,
\end{equation}
where the node index $j$ is randomly chosen from $\{1,\cdots,N\}$. It is straightforward to prove that $\rho=\rho_{kx}$~\cite{Eom2014}. $\rho$ can have a value in $[-1,1]$. The attribute has the average $\langle x\rangle=(\rho+\sqrt{1-\rho^2})\langle k\rangle$, while its standard deviation is the same as that of degrees, i.e., $\sigma_x=\sigma_k$, independent of $\rho$. From the generated attribute sequence, one can measure the attribute-attribute correlation $r_{xx}$ using Eq.~(\ref{eq:assort}) but with $k$ replaced by $x$. $r_{xx}$ can be interpreted as the degree of attribute homophily~\cite{McPherson2001}. For comparison to the analytic solution in Eq.~(\ref{eq:Pkx_gamma}), we assume the gamma distribution for the degree as in Eq.~(\ref{eq:gammaDistr}). Since the analytic results are not sensitive to the variation of $\alpha$, we use $\alpha=1$ for simplicity. The other form of degree distribution, e.g., power-law distribution, has been studied in Appendix.

\begin{figure}[!t]
    \includegraphics[width=.95\columnwidth]{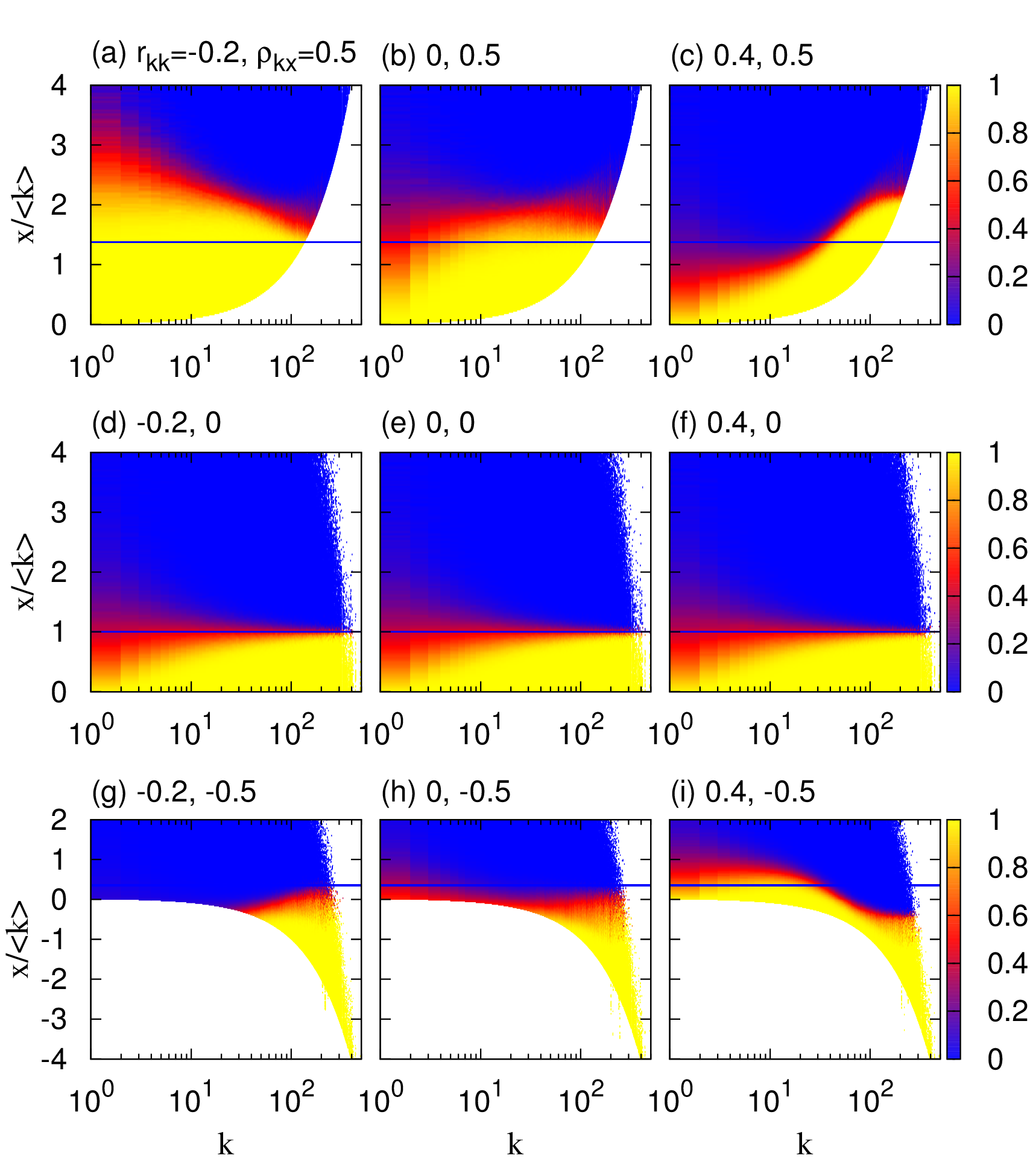}
    \caption{(Color online) Paradox holding probability $h_k(x)$ of the correlated networks of size $N=10^5$ for values of $r_{kk}=-0.2$, $0$, and $0.4$ (from left to right) and of $\rho_{kx}=-0.5$, $0$, and $0.5$ (from bottom to top). Degrees $k$ follow the gamma distribution in Eq.~(\ref{eq:gammaDistr}) with $\alpha=1$ and $\beta=50$, i.e., $\langle k\rangle=50$, and attributes $x$ are assigned to nodes using Eq.~(\ref{eq:assign_x}). For comparison to the uncorrelated case, $x$ has been regularized by $\langle k\rangle$ that has the same value as $\langle x\rangle$ for $\rho_{kx}=0$. Blue horizontal lines correspond to $\langle x\rangle/\langle k\rangle$ for each case.}
    \label{fig:gamma_correl}
\end{figure}

Let us first consider the uncorrelated case, i.e., $r_{kk}=\rho_{kx}=0$. We generate an uncorrelated network of size $N=10^5$ and of $\langle k\rangle=\langle x\rangle=50$. Then we measure the paradox holding probability $h_k(x)$ to find that the numerical result in Fig.~\ref{fig:gamma_correl}(e) supports our analytic solution of Eq.~(\ref{eq:Pkx_gamma}), also depicted in Fig.~\ref{fig:gamma_uncorrel}(a). The values of $h_k(x)$ for $x/\langle x\rangle=0.9$, $1$, and $1.1$ are plotted in Fig.~\ref{fig:gamma_uncorrel}(b) for the precise comparison to the analytic solution. In all cases, $h_k(x)$ has been averaged over $100$ different assignments of attributes using Eq.~(\ref{eq:assign_x}).

In general, the paradox holding probability is expected to be affected by the combined effect of two correlations, i.e., $r_{kk}$ and $\rho_{kx}$. As shown in Fig.~\ref{fig:gamma_correl}(d--f), when $\rho_{kx}=0$, the overall behavior of $h_k(x)$ is the same as the uncorrelated case in Fig.~\ref{fig:gamma_uncorrel}(a), irrespective of $r_{kk}$. It is because attributes of neighboring nodes are fully uncorrelated, supported by the observation of $r_{xx}\approx 0$. By the same argument, the similar pattern is observed for $r_{kk}=0$ and $\rho_{kx}\neq 0$. This is evidenced by the fact that the border $x_k$, defined by the condition $h_k(x=x_k)=1/2$, is mostly flat for a wide range of $k$. However, such borders show some deviations from $x=\langle x\rangle$, depicted by blue horizontal lines in Fig.~\ref{fig:gamma_correl}, possibly due to finite size effects.

\begin{figure}[!t]
    \includegraphics[width=.95\columnwidth]{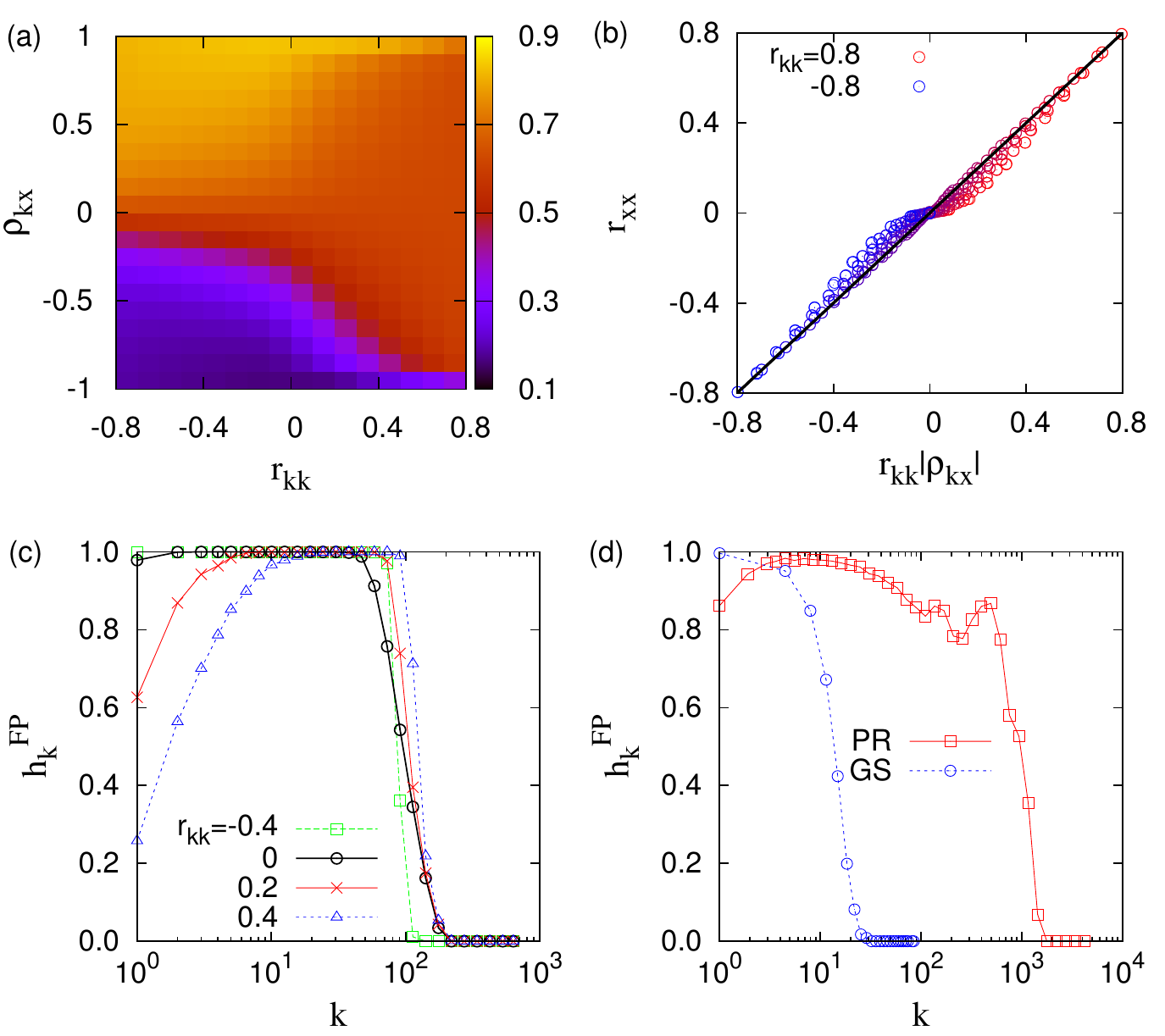}
    \caption{(Color online) Numerical results for correlated networks of size $N=10^5$ and of $\langle k\rangle=50$ with the gamma distribution for degrees (a--c): (a) Average paradox holding probability $H$ as a function of $r_{kk}$ and $\rho_{kx}$. (b) Scatter plot showing $r_{xx}$ and $r_{kk}|\rho_{kx}|$ for $-0.8\leq r_{kk}\leq 0.8$. The solid line corresponds to $r_{xx}=r_{kk}|\rho_{kx}|$. (c) Paradox holding probability of the FP for various values of degree-degree correlations. (d) Empirical paradox holding probability of the FP for coauthorship networks of Physical Review journals (PR) and Google Scholar profiles (GS) from~\cite{Eom2014}.}
    \label{fig:gamma_correl_summary}
\end{figure}

When both $r_{kk}$ and $\rho_{kx}$ are positive [Fig.~\ref{fig:gamma_correl}(c)], the effect of attribute homophily by $r_{xx}>0$ becomes pervasive. The GFP holds for high attribute nodes due to their neighbors of even higher attributes, while low attribute nodes have lower paradox holding probability, compared to the uncorrelated case. The opposite behavior is observed for the dissortative networks [Fig.~\ref{fig:gamma_correl}(a)]. Hub nodes of high attribute tend to be connected with dangling nodes of low attribute, leading to smaller $h_k(x)$ for the former and larger $h_k(x)$ for the latter. It also means the negative attribute-attribute correlation ($r_{xx}<0$). Let us now consider when degrees and attributes are negatively correlated ($\rho_{kx}<0$). In the assortative networks [Fig.~\ref{fig:gamma_correl}(i)], the GFP holds even for some high attribute nodes but with small degrees, which is comparable to the case of $r_{kk}, \rho_{kx}>0$. In the dissortative networks [Fig.~\ref{fig:gamma_correl}(g)], hub nodes of low attribute tend to be connected to dangling nodes of high attribute, leading to larger $h_k(x)$ for the former and smaller $h_k(x)$ for the latter. This is in contrast to the case of $r_{kk}<0$ and $\rho_{kx}>0$. It is notable that the results for $r_{xx}\approx 0$ and for $r_{kk}, \rho_{kx}>0$ are comparable to empirical results for coauthorship networks of PR and GS in Fig.~1(d,f) and Fig.~1(a,c) of~\cite{Eom2014}, respectively.

Now we calculate the average paradox holding probability $H(r_{kk},\rho_{kx})$, which is defined as the fraction of nodes satisfying Eq.~(\ref{eq:individualGFP}). The result is shown in Fig.~\ref{fig:gamma_correl_summary}(a). As a reference, we define $H_0\equiv H(0,0)\approx 0.62$ for the uncorrelated case. If $r_{kk}\lesssim 0.4$, it is found that $H>H_0$ ($H<H_0$) for $\rho_{kx}>0$ ($\rho_{kx}<0$). Otherwise, if $r_{kk}>0.4$, $H\approx H_0$ is observed for almost entire range of $\rho_{kx}$. We first note that most nodes in the network have small degrees from the gamma distribution, and they have low attributes if $\rho_{kx}\geq 0$ or high attributes but around $0$ for $\rho_{kx}<0$. These nodes dominate the population, hence the behavior of $H$. Next, the paradox holding probability of such dominant nodes needs to be understood. In the dissortative networks ($r_{kk}<0$), large degree nodes tend to be connected to small degree nodes, leading to a star-like structure. If hub nodes have high attributes and peripheral nodes have low attributes ($\rho_{kx}>0$), the dominant nodes, i.e., peripheral nodes in this case, have large paradox holding probability, resulting in $H>H_0$. Otherwise, if $\rho_{kx}<0$, since the dominant nodes have high attribute, we find $H<H_0$. Here the attributes of neighboring nodes are negatively correlated ($r_{xx}<0$) irrespective of the sign of $\rho_{kx}$. In the assortative networks ($r_{kk}>0$), nodes of similar degrees tend to be connected to each other. The attributes of neighboring nodes are similar ($r_{xx}>0$) whether high (low) degree nodes have high (low) attributes ($\rho_{kx}>0$) or vice versa ($\rho_{kx}<0$). In either case, the dominant nodes have neighbors of similar attribute, implying that the behavior of $H$ is robust against the variation and sign of $\rho_{kx}$. Conclusively, the sign of $\rho_{kx}$ is relevant to $H$ in the dissortative network with $r_{kk}<0$, while it is irrelevant to $H$ in the assortative network with $r_{kk}>0$. This can be compared to the GFP at the network level, which is determined by the sign of $\rho_{kx}$ as shown in Eq.~(\ref{eq:networkGFP}). We also numerically find that $r_{xx}\approx r_{kk}|\rho_{kx}|$ in Fig.~\ref{fig:gamma_correl_summary}(b), implying that the behavior of $H$ cannot be explained only in terms of $r_{xx}$.

Finally, using the above generated networks, we calculate the probability of holding the FP, denoted by $h_k^{\rm FP}$. As shown in Fig.~\ref{fig:gamma_correl_summary}(c), for $r_{kk}\leq 0$, $h_k^{\rm FP}$ stays close to $1$ until $k$ reaches $\approx 100$, and decays quickly to $0$. It is because small degree nodes tend to be connected to large degree nodes. 
However, in the assortative networks with $r_{kk}>0$, $h_k^{\rm FP}$ begins with small values, increases according to $k$, and eventually decays to $0$. It implies that the FP holds most strongly for nodes of average degree, or so-called middle class, not for nodes of the smallest degree. These variations at the individual level are observed only due to different effects of assortativity coefficient, $r_{kk}$. In contrast, the FP at the network level is influenced only by the shape of degree distribution, irrespective of $r_{kk}$.
These results enable us to understand the empirical finding of $h_k^{\rm FP}$ from coauthorship networks~\cite{Eom2014}, replotted in Fig.~\ref{fig:gamma_correl_summary}(d). The increasing behavior of $h_k^{\rm FP}$ for $k<10$ in the coauthorship network of PR is due to $r_{kk}\approx 0.47$, while such increasing behavior is not observed in the coauthorship network of GS showing no degree-degree correlation, i.e., $r_{kk}\approx -0.02$.

\begin{figure}[!t]
    \includegraphics[width=.95\columnwidth]{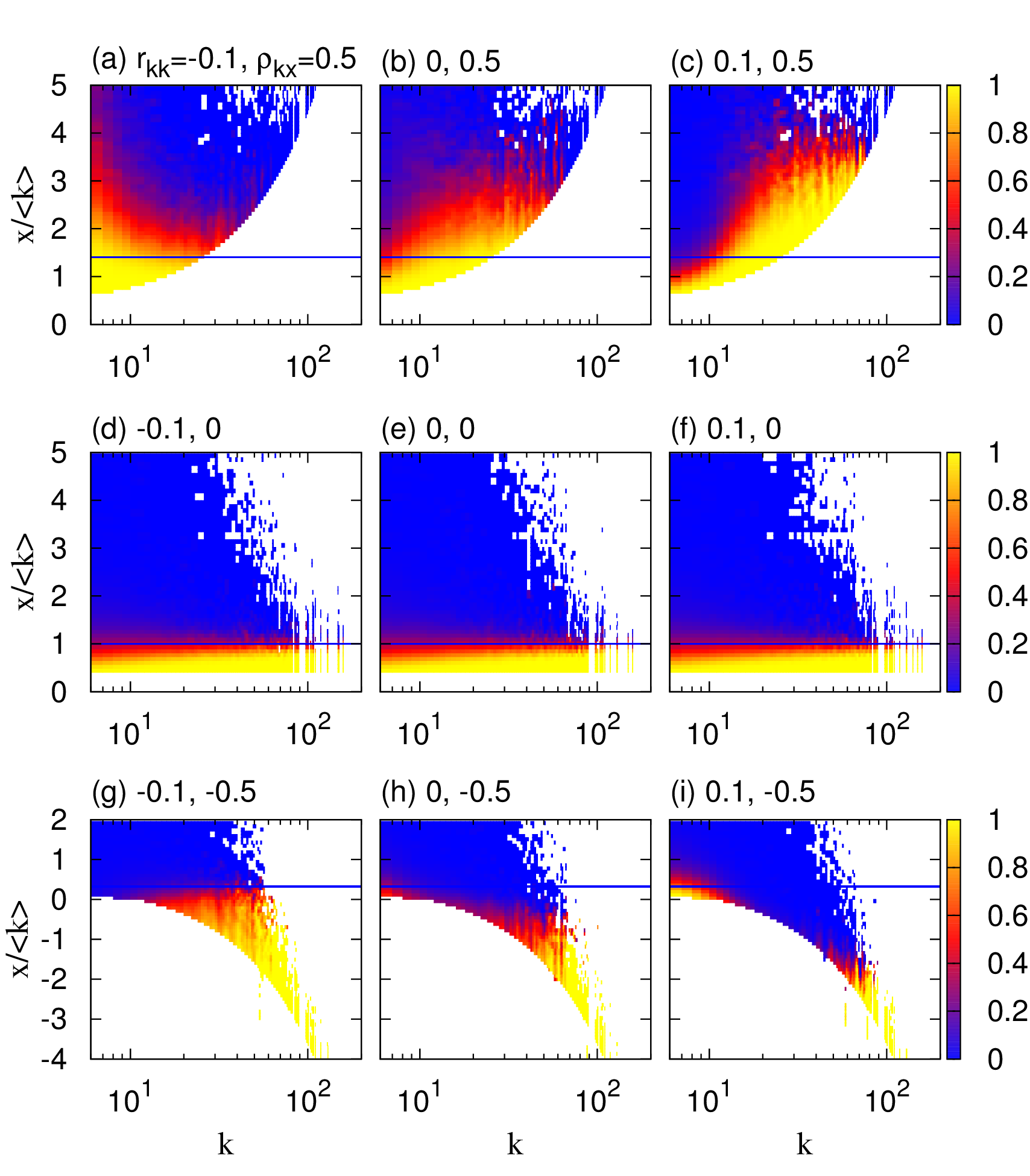}
    \caption{(Color online) Paradox holding probability $h_k(x)$ of the correlated networks of size $N=10^4$ for values of $r_{kk}=-0.1$, $0$, and $0.1$ (from left to right) and of $\rho_{kx}=-0.5$, $0$, and $0.5$ (from bottom to top). Degrees $k$ follow the power-law distribution in Eq.~(\ref{eq:powerDistr}) with $\gamma=2.7$ and $k_{\rm min}=6$, and attributes $x$ are assigned to nodes using Eq.~(\ref{eq:assign_x}). For comparison to the uncorrelated case, $x$ has been regularized by $\langle k\rangle$ that has the same value as $\langle x\rangle$ for $\rho_{kx}=0$. Blue horizontal lines correspond to $\langle x\rangle/\langle k\rangle$ for each case.}
    \label{fig:power_correl}
\end{figure}

\begin{figure}[!t]
    \includegraphics[width=.95\columnwidth]{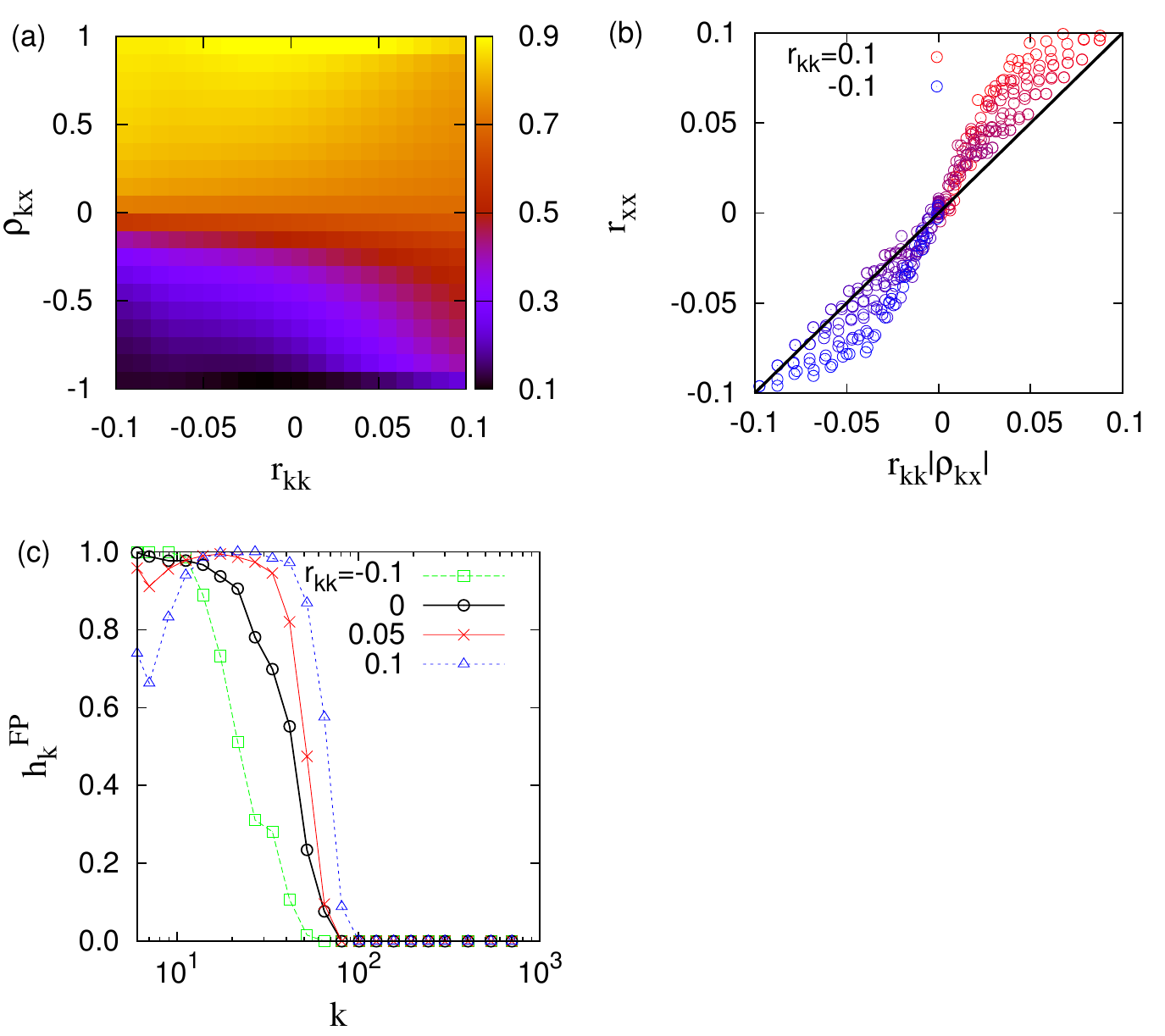}
    \caption{(Color online) Numerical results for correlated networks of size $N=10^4$ with the power-law distribution for degrees: (a) Average paradox holding probability $H$ as a function of $r_{kk}$ and $\rho_{kx}$. (b) Scatter plot showing $r_{xx}$ and $r_{kk}|\rho_{kx}|$ for $-0.1\leq r_{kk}\leq 0.1$. The solid line corresponds to $r_{xx}=r_{kk}|\rho_{kx}|$. (c) Paradox holding probability of the FP for various values of degree-degree correlations.}
    \label{fig:power_correl_summary}
\end{figure}

\section{Conclusions}

As an interplay between topological heterogeneities and node attributes in complex networks, the generalized friendship paradox (GFP) has been recently suggested, implying that your friends have on average higher attribute than you do~\cite{Eom2014}. While the GFP at the network level was clearly explained in terms of the positive degree-attribute correlations, the GFP at the individual level has been far from being fully understood. In order to understand the role of degree-attribute correlations for the GFP at the individual level in more detail, we analyze the uncorrelated solvable model, which already shows nontrivial behavior especially for high attribute nodes. For the general case, we numerically study the correlated network model with tunable degree-degree and degree-attribute correlations, denoted by $r_{kk}$ and $\rho_{kx}$, respectively. We obtain the detailed patterns of the paradox holding probability of individuals depending on their degrees and attributes, for the entire range of correlations of $r_{kk}$ and $\rho_{kx}$. Similarly to the GFP at the network level, the average paradox holding probability is strongly affected by the sign of $\rho_{kx}$ only in the dissortative networks with $r_{kk}<0$. On the other hand, the results for the assortative networks with $r_{kk}>0$ are robust against the variation and sign of $\rho_{kx}$. 

In our study, we have ignored other topological heterogeneities of networks like community structure~\cite{Fortunato2010}, and assumed that node attributes are fixed and do not change. As future works, it would be interesting to study the GFP in more realistic network topology and/or in case where the attributes can change in time such as the attractiveness of scientific papers~\cite{Eom2011}, or they evolve according to the individual decisions, e.g., within the framework of evolutionary game theory~\cite{Hauert2005}.

Finally, we like to remark that successful applications of statistical physics to social phenomena necessitate the detailed understanding of both objective and subjective sides of individual behaviors. In this sense, our study of the GFP can provide insights for the subjective self-evaluation of individuals compared to their neighbors~\cite{Zuckerman2001,Kross2013}, which shapes the way how they interact with others. This is crucial to understand the emergent collective dynamics at the community or societal level.

\begin{acknowledgements}
We gratefully acknowledge the Aalto University postdoctoral program (H.-H.J.) and the EC FET Open project ``New tools and algorithms for directed network analysis,'' NADINE number 288956 (Y.-H.E.) for financial support.
\end{acknowledgements}

\appendix*
\section{Correlated model with power-law distribution}

We study the GFP for the correlated networks with tunable degree-attribute correlations for the power-law distribution of degrees and attributes. In case of power-law degree distribution, the degree-degree correlation $r_{kk}$ is strongly limited by various factors like the system size and the power-law exponent of degree distribution, as studied in~\cite{Menche2010}. For the realistic consideration, we choose the following distribution
\begin{equation}
  \label{eq:powerDistr}
  P(k)\propto k^{-\gamma}\ \textrm{for}\ k\geq k_{\rm min},
\end{equation}
with $\gamma=2.7$ and $k_{\rm min}=6$. For these values of parameters, one can generate the network in the range of $-0.1\leq r_{kk}\leq 0.1$ for $N=10^4$. Then, we calculate the paradox holding probability $h_k(x)$ to find that its overall behavior is qualitatively similar to those in the case of gamma distribution, as shown in Fig.~\ref{fig:power_correl}. We also find the similar behaviors for average paradox holding probability $H(r_{kk},\rho_{kx})$, for the linear relationship between $r_{xx}$ and $r_{kk}|\rho_{kx}|$ but with larger deviations due to the relatively narrow range of $r_{kk}$, and for the probability of holding the FP for various values of degree-degree correlation. The results are summarized in Fig.~\ref{fig:power_correl_summary}.


\end{document}